\documentclass[aps,pre,twocolumn,superscriptaddress,showkeys]{revtex4-1}
\usepackage{amsmath,amssymb,amsfonts,amsthm,enumerate,multirow}
\usepackage{natbib}
\usepackage{graphicx}

\begin{document}

\title{Deformation and Fabric in Compacted Clay Soils}

\author{CM Wensrich}
\email[]{Christopher.Wensrich@newcastle.edu.au}
\affiliation{School of Engineering, The University of Newcastle, Australia}

\author{J Pineda}
\affiliation{ARC Center of Excellence for Geotechnical Science and Engineering, The University of Newcastle, Australia}

\author{V Luzin}
\affiliation{Australian Centre for Neutron Scattering, Australian Nuclear Science and Technology Organisation, Australia}

\author{L Suwal}
\affiliation{ARC Center of Excellence for Geotechnical Science and Engineering, The University of Newcastle, Australia}

\author{EH Kisi}
\affiliation{School of Engineering, The University of Newcastle, Australia}

\author{H Allameh-Haery}
\affiliation{School of Engineering, The University of Newcastle, Australia}

\date{\today}

\begin{abstract}

Hydro-mechanical anisotropy of clay soils in response to deformation or deposition history is related to the micromechanics of plate-like clay particles and their orientations.  In this letter, we examine the relationship between microstructure, deformation and moisture content in kaolin clay using a technique based on neutron scattering.  This technique allows for the direct characterisation of microstructure within representative samples using traditional measures such as orientation density and soil fabric tensor.  From this information, evidence for a simple relationship between components of the deviatoric strain tensor and the deviatoric fabric tensor emerged.  This relationship may provide a physical basis for future anisotropic constitutive models based on the micromechanics of these materials.

\end{abstract}



\maketitle

Compacted clays are found in a wide variety of natural and man-made settings.  They show complex hydro-mechanical behaviour that must be understood for effective design of civil infrastructure such as dams, embankments, roads as well as nuclear waste disposal facilities. 

\begin{figure*} [!htb]
    	\centering
    	\includegraphics[height=0.21\linewidth]{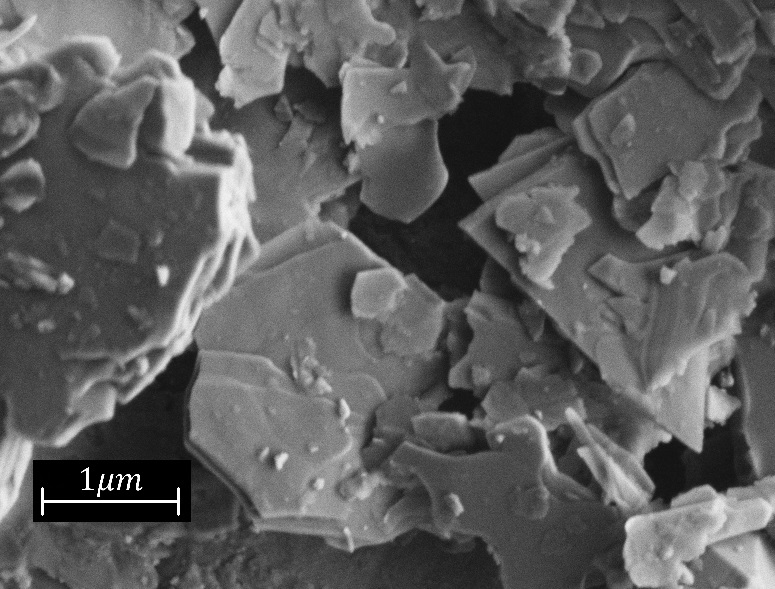}
    	\put(-140,115){(a)}
    	\hspace{5ex}
       	\includegraphics[height=0.22\linewidth]{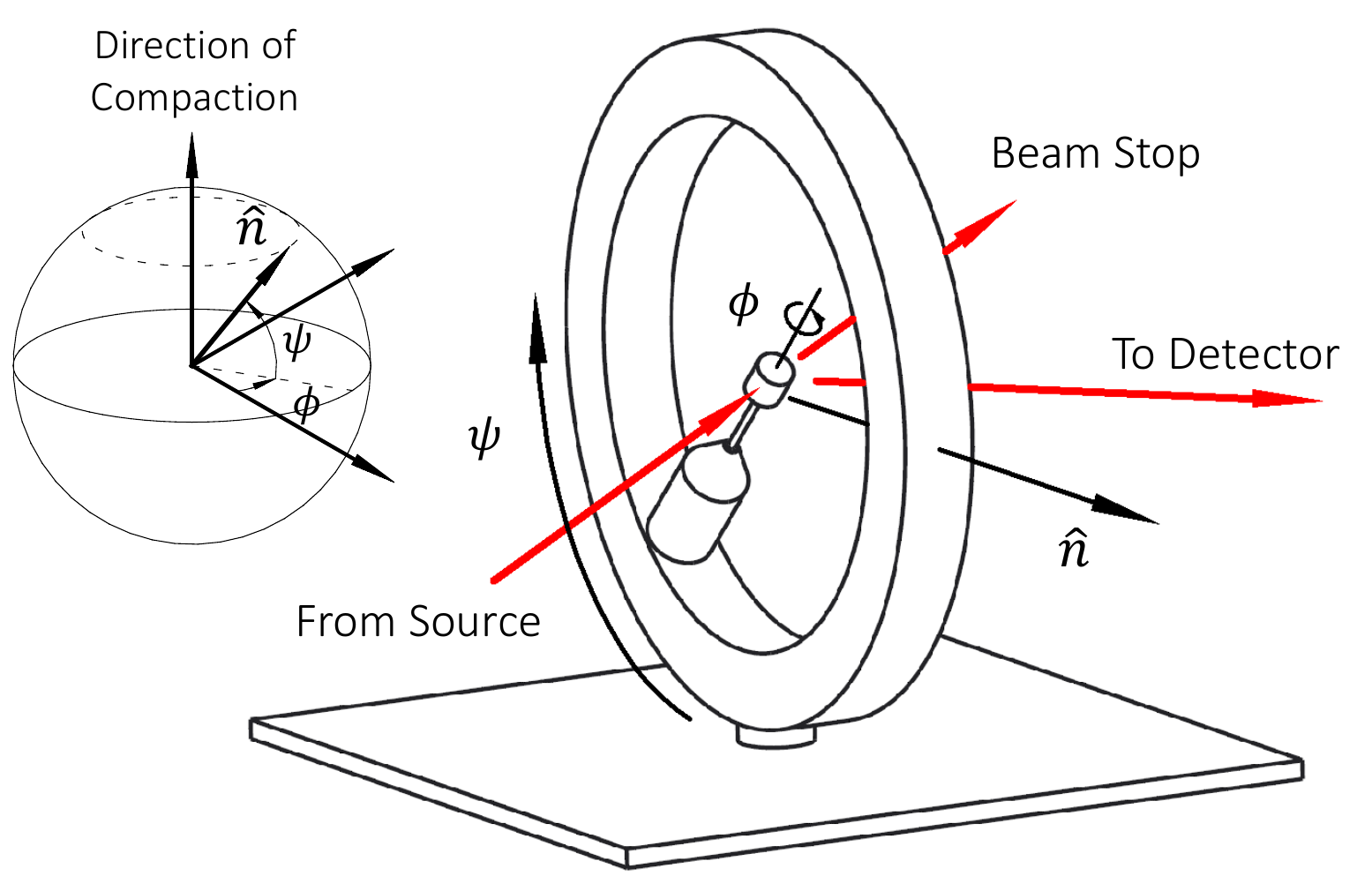}
       	\put(-180,115){(b)}
       	\hspace{5ex}
	\includegraphics[height=0.22\linewidth]{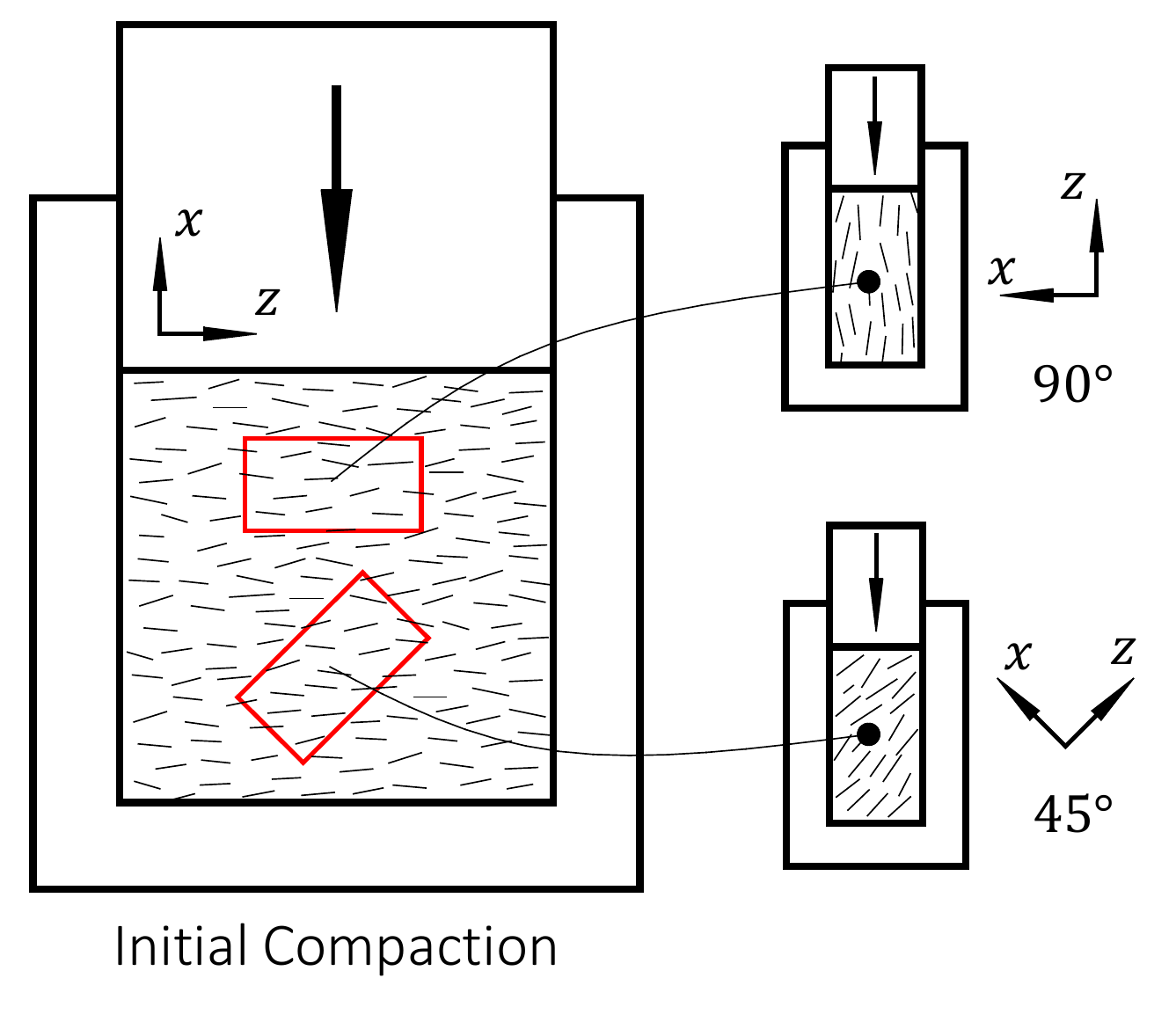}
	\put(-140,115){(c)}
	\caption{(a) SEM image of plate-like kaolin particles.  (b) Texture measurement on a neutron instrument.  The intensity of the diffracted beam is proportional to the number of crystal  planes aligned in the direction $\hat{n}$. (c) Sample preparation for the second experiment; initial compaction to a reference state followed by compaction at either 90$^\circ$ or 45$^\circ$ to the initial direction.}
    	\label{fig:ExpSetup}
    	\centering
\end{figure*}

One of the more complicated aspects of these materials is the significant amount of anisotropy they can display. This anisotropy is directly linked to micro-structure; properties are sensitive to  preferred orientations of the microscopic plate-like clay particles (see Figure \ref{fig:ExpSetup}a).  The arrangement of these platelets is often referred to as the \textit{fabric} of the material, the evolution of which is directly related to deformation and deposition history. Modern constitutive models strive to capture these effects by modifying (rotating) the yield surface and flow rules in response to plastic strain \cite{wheeler2003anisotropic,voyiadjis2000finite}. This approach assumes a causal link between deformation, fabric and mechanical anisotropy, however the micro-mechanics of this are poorly understood.  

The development of a proper understanding of the mechanisms behind anisotropy in clays relies on being able to accurately characterise micro-structure.  Since the pioneering work of Barden \textit{et al.} \cite{barden1973collapse}, Collins \textit{et al.} \cite{collins1974form}, Osipov \textit{et al.} \cite{osipov1978structure} and Sloane and Kell \cite{sloane1966fabric}, various qualitative and quantitative techniques have been combined to approach this goal.  Mercury Intrusion Porosimetry (MIP) and nitrogen adsorption tests \cite{diamond1970pore,romero2008microstructure} can examine pore size distribution, however fabric represents a greater challenge.  Traditional techniques involved microscopy (SEM and ESEM) and laboratory-based X-ray sources (e.g. \cite{montes2005esem,lin2014applications,suuronen2014x,haines2009clay}), however these methods can be limited to small samples or observations near external surfaces which are not necessarily representative.  

More recent techniques using synchrotron and neutron radiation \cite{schumann2014texture,voltolini2008anisotropy,cifelli2005origin} can explore bulk properties. The principle involved is as follows;

The plate-like nature of clay particles is directly related to their layered silicate crystal structure.  As a consequence, the physical orientation of a clay platelet corresponds to crystallographic directions.  In other words, the crystallographic \textit{texture} of a clay sample directly informs on micro-structural fabric.  Crystallographic texture, expressed as the orientation density of crystal planes, is of prime concern in various areas of material science and a number of experimental techniques based on diffraction are available to make such measurements.  In particular, the penetrating nature of neutrons provides a method by which texture can be measured in bulk samples at the scale of centimeters depending on the material \cite{wenk2006neutron}.   The approach relies upon the fact that the intensity of a diffracted beam depends directly on the number of crystal grains that satisfy Bragg's law for a given orientation of the sample and instrument.  Figure \ref{fig:ExpSetup}b provides a typical experimental setup.

With an incident beam of constant wavelength, the magnitude of a diffraction peak, $I_{hkl}$, as a function of direction (relative to the sample), $\boldsymbol{\hat{n}} = (\cos \psi \cos \phi, \cos \psi \sin \phi, \sin \psi)$, can be used to map-out the orientation density for the corresponding lattice planes;
\begin{equation}
	O_{hkl}(\boldsymbol{\hat{n}})=\frac{I_{hkl}(\boldsymbol{\hat{n}})}{\langle I_{hkl} \rangle}
\end{equation}

In the case of a kaolinite, the orientation density of the basal lattice planes (e.g. \textit{0001}, \textit{0002} etc.) directly provide the orientation density of platelets.  Note that there is a slight misuse of terminology here.  The term ``orientation'' usually refers to the directions of all lattice planes (i.e. as specified by 3 Euler angles).  In our case, orientation of platelets about their normal axis is irrelevant (and most likely random) as far as fabric is concerned.  For this reason, in this letter our use of the term ``orientation' is interchangeable with ``direction' or ``axis''.

Orientation density can be used to calculate fabric tensors of the form \cite{ken1984distribution};
\begin{equation}
	F_{i_1i_2 \cdots i_r}=\tfrac{1}{2 \pi} \int_0^{\tfrac{\pi}{2}} \int_0^{2\pi} O(\boldsymbol{\hat{n}}) \hat{n}_{i_1} \hat{n}_{i_2}\cdots\hat{n}_{i_r} \cos \psi  d\phi d\psi,
\end{equation}
where $r$ is an arbitrary rank (note that $O(\boldsymbol{\hat{n}})=O(-\boldsymbol{\hat{n}})$).  In each case this represents an ensemble average of the form $F_{i_1i_2 \cdots i_r}=\langle \hat{h}_{i_1} \hat{h}_{i_2}\cdots\hat{h}_{i_r} \rangle$, over individual platelet unit normal vectors, $\boldsymbol{\hat{h}}$, within the sample.

Fabric tensors of this type are often used in soil mechanics as a description of internal structure.  In granular materials, a fabric tensor usually refers to orientations of contacts, however the plate-like structure of clay particles implies equivalence in this case.  Higher rank tensors can capture higher order variations, however in practice, rank-2 is usually deemed sufficient.  The trace of this tensor is unity, with a random distribution of orientations implied by $F_{ij}=\tfrac{1}{3}\delta_{ij}$.  In principal directions, diagonal components of the rank-2 fabric tensor refer to the relative size of principal diameters of an ellipsoidal orientation density function - higher rank tensors represent higher order spherical harmonics.  

In this letter we utilise neutron scattering techniques in two experiments aimed at examining the evolution of fabric within kaolin-clay samples as a function of composition, density and deformation history. 

The first experiment involved an examination of the evolution of soil fabric during compaction along with the effects of moisture content;

Clay samples were prepared by mixing kaolin powder with deionized water before sieving to produce a maximum aggregate size of 1.5 mm. After an equalization time of 48 hours, eight cylindrical samples ($\varnothing 15 \times 10$mm) were uniaxially compacted to the densities and moisture contents given in Table \ref{tab:samples1}. After compaction, all moisture was removed through the freeze-drying method \cite{delage1982use}.

 \begin{table}
 \caption{\label{tab:samples1} Sample properties for the first experiment}
 \begin{ruledtabular}
 \begin{tabular}{c | c c }
 
 Sample & Dry  & Moisture Content\\
 Number & Density [g/cc]&  (by mass) \\
 \hline
 S1-0 & 1.85 & 8.17\% \\
 S2-0 & 1.70 & 8.17\% \\
 S3-0 & 1.54 & 8.17\% \\
 S4-0 & 1.32 & 8.17\% \\
 S5-0 & 1.30 & 2.4\% \\
 S6-0 & 1.30 & 18.3\% \\
 S7-0 & 1.32 & 27.5\% \\
 S8-0 & 1.31 & 33\% \\
 
 \end{tabular}
 \end{ruledtabular}
 \end{table}

Orientation density within each sample was then measured using the KOWARI diffractometer at the Australian Centre for Neutron Scattering (ACNS) within the Australian Nuclear Science and Technology Organisation (ANSTO).  These measurements were based on the relative intensity of the (0002) diffraction peak from monochromatic neutrons of wavelength 2.8\AA over a $5^\circ \times 5^\circ$ regular grid of sample orientations for $0<\phi<360$ and $0<\psi<90$ using a standard 4-circle goniometer.  High background levels due to hydrogen within hydroxyl groups necessitated relatively long sampling times; each sample required a total of around 12 hours of beamtime.

\begin{figure*} [!htb]
    	\centering
       	\includegraphics[width=0.55\linewidth]{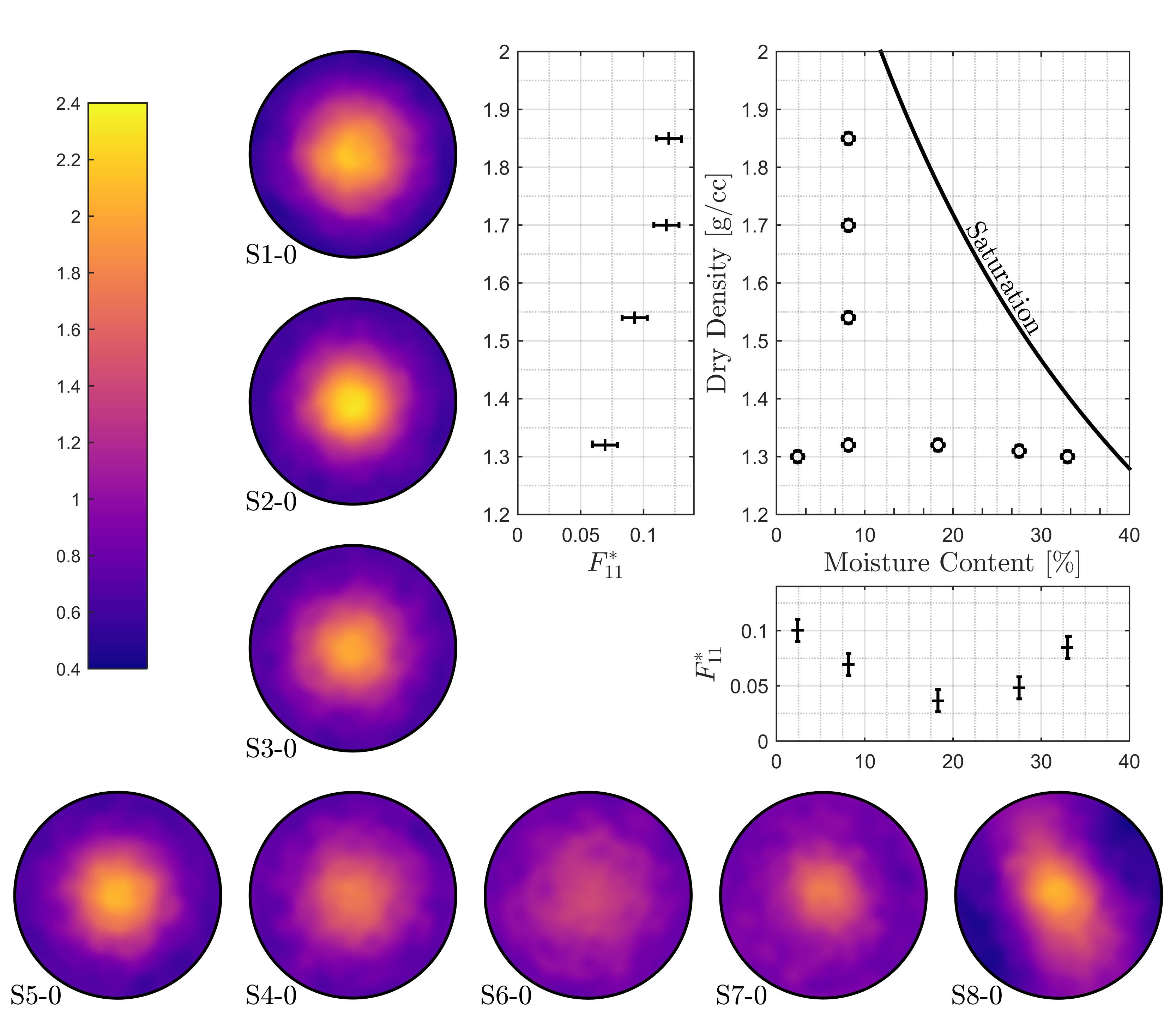}
       	\put(-280,230){(a)}
 	\includegraphics[width=0.4\linewidth]{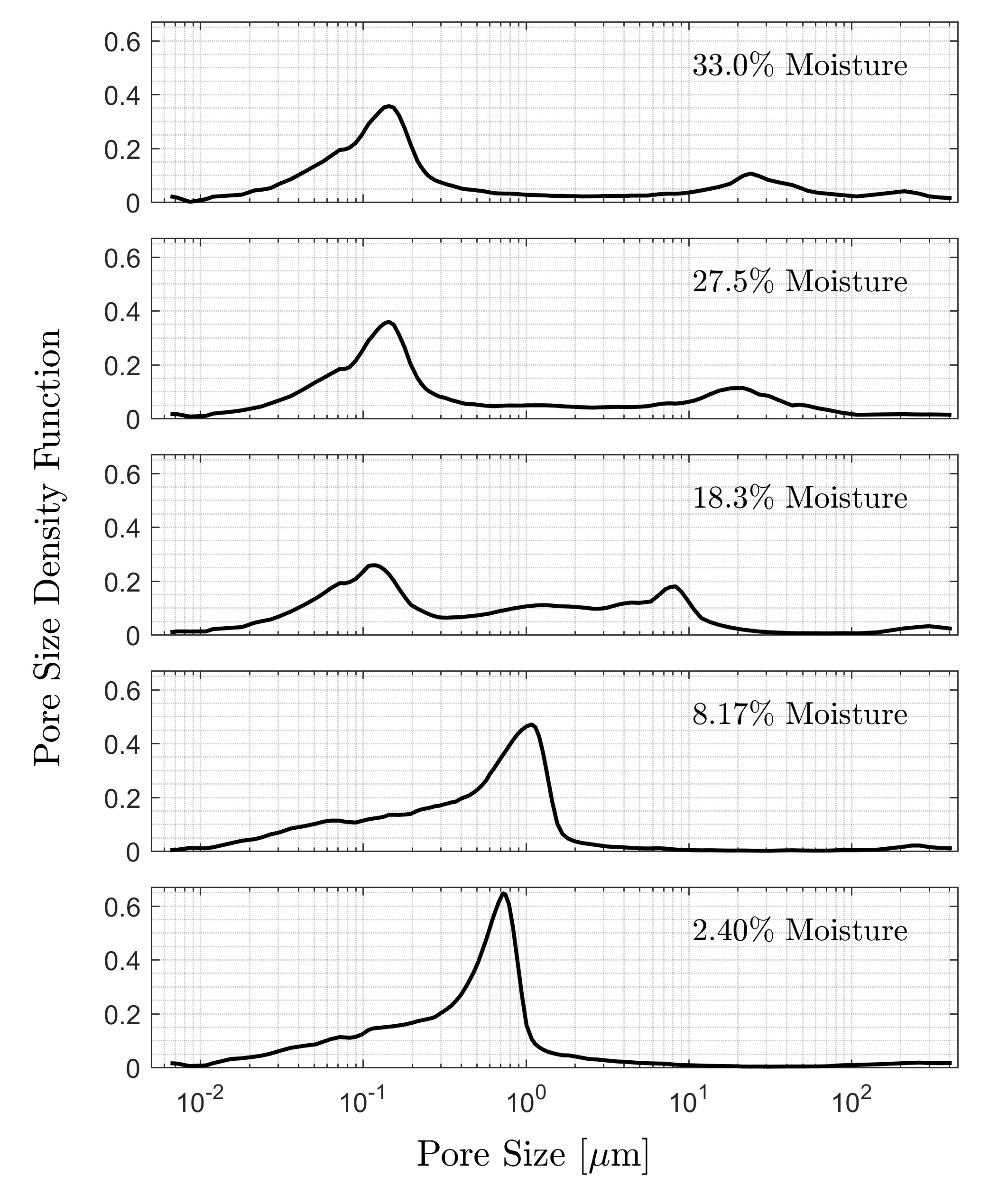}
 	\put(-200,230){(b)}
	\caption{(colour online) Results showing the evolution of clay fabric during compaction and the effects of moisture content.  (a) Pole figures show the 3D orientation density function for platelets as viewed in the direction of compaction. (b) Pore size distribution from MIP as a function of moisture content for a series of samples with 1.3g/cc dry density.}
    	\label{fig:Exp1}
    	\centering
\end{figure*}

Figure \ref{fig:Exp1}a shows the results of this experiment.  8 pole figures are shown arranged in the order indicated in the graph in the upper right corner.  Each pole figure is a depiction of the orientation density as viewed along the direction of compaction (notionally aligned with the third coordinate axis).  Also shown in the upper right corner is the $11$-component of the deviatoric part of the rank-2 fabric tensor, $F^*_{ij}=F_{ij}-\tfrac{1}{3}\delta_{ij}$, plotted as a function of either final density or moisture content.  In this case, the $F_{11}$ component provides a measure of the degree of alignment of platelets to the direction of compaction.

With the exception of the wettest sample (S8-0), all of the pole figures indicate \textit{fibre textures} aligned with the direction of compaction.  As expected, this indicates an alignment of platelets with their axis normal to the direction of compaction.  The wettest sample shows a similar orientation density superimposed with a faint meridional band.  This is thought to be due to difficulties in the preparation of this sample associated with large agglomerates present prior to compaction; it is likely that deformation of these agglomerates was not strictly axial.

Overall, the measurements show two interesting trends.  First, the deviation of the fabric from the isotropic state of $F^*=0$ increases with the level of compaction. This result is perhaps expected.  Moisture seems to have a more varied effect.  For the same density, low and high levels of moisture lead to higher levels of alignment with a minimum in-between.  To explore this behaviour, the pore structure of a second series of samples compacted to the same density was examined using MIP.  Figure \ref{fig:Exp1}b shows the Pore Size Distributions (PSD) of these samples arranged in decreasing moisture content.  Specimens compacted at low moisture contents (less than 10\%) clearly show a mono-modal PSD.  A mono-modal PSD is also predominant at higher moisture contents although the presence of some macro-porosity is also recognized.  In between (particularly at 18.3\% moisture) a clear bi-modal PSD can be observed due to clay aggregation. Bi-modal PSDs are commonly related to random (open) fabrics \cite{burton2015microstructural}. It is likely that the deformation of the aggregates is not strictly axial, leading to the minimum value in $F^*_{11}$ observed in Figure \ref{fig:Exp1}a. 

Following the success of this experiment, a second experiment was devised to examine the relationship between the evolution of fabric and deformation for more complex loading paths.  This experiment involved the preparation of samples through the following process;

\begin{figure*}
    	\centering
    	\includegraphics[width=0.40\linewidth]{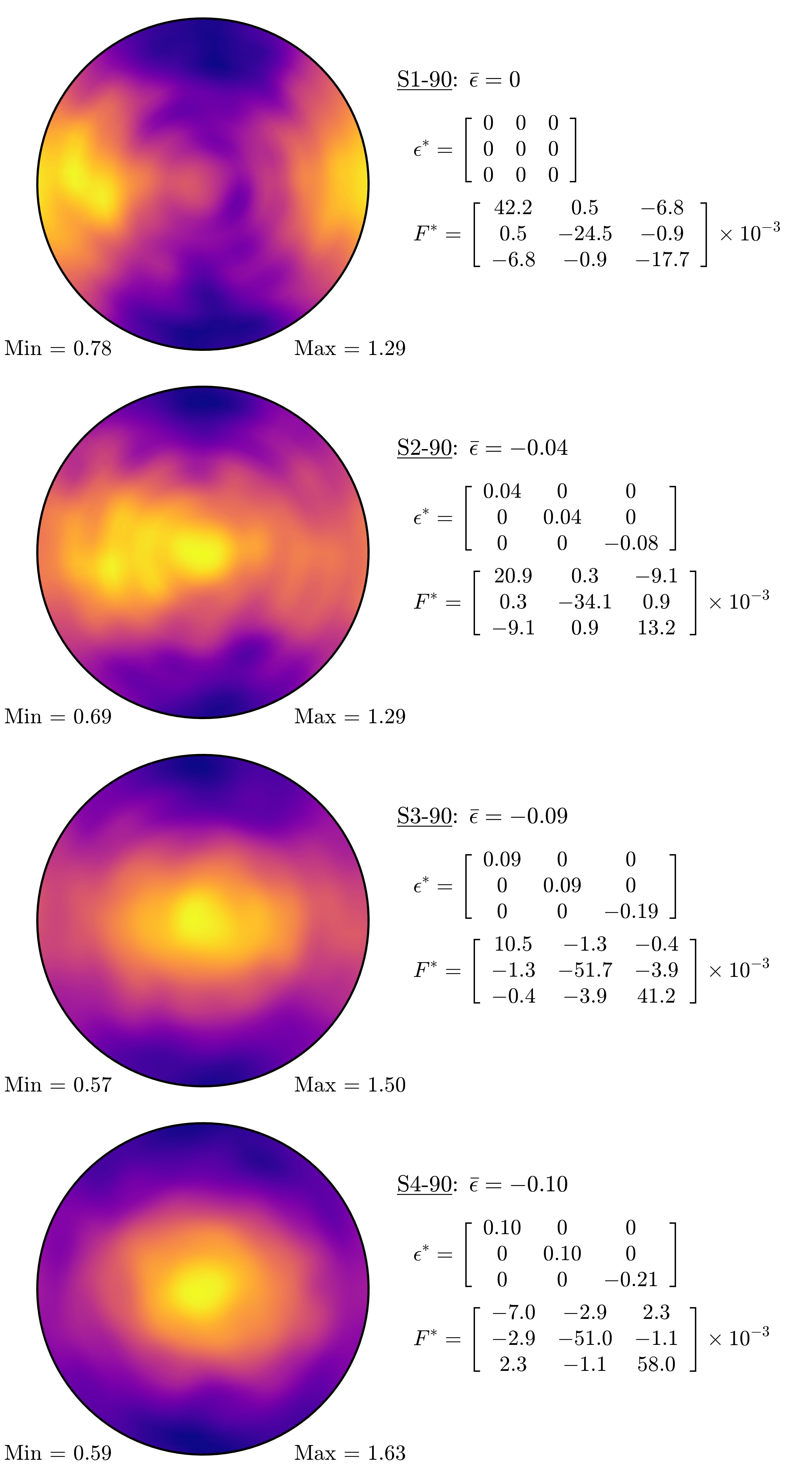}
    	\put(-200,360){(a)}
       	\includegraphics[width=0.40\linewidth]{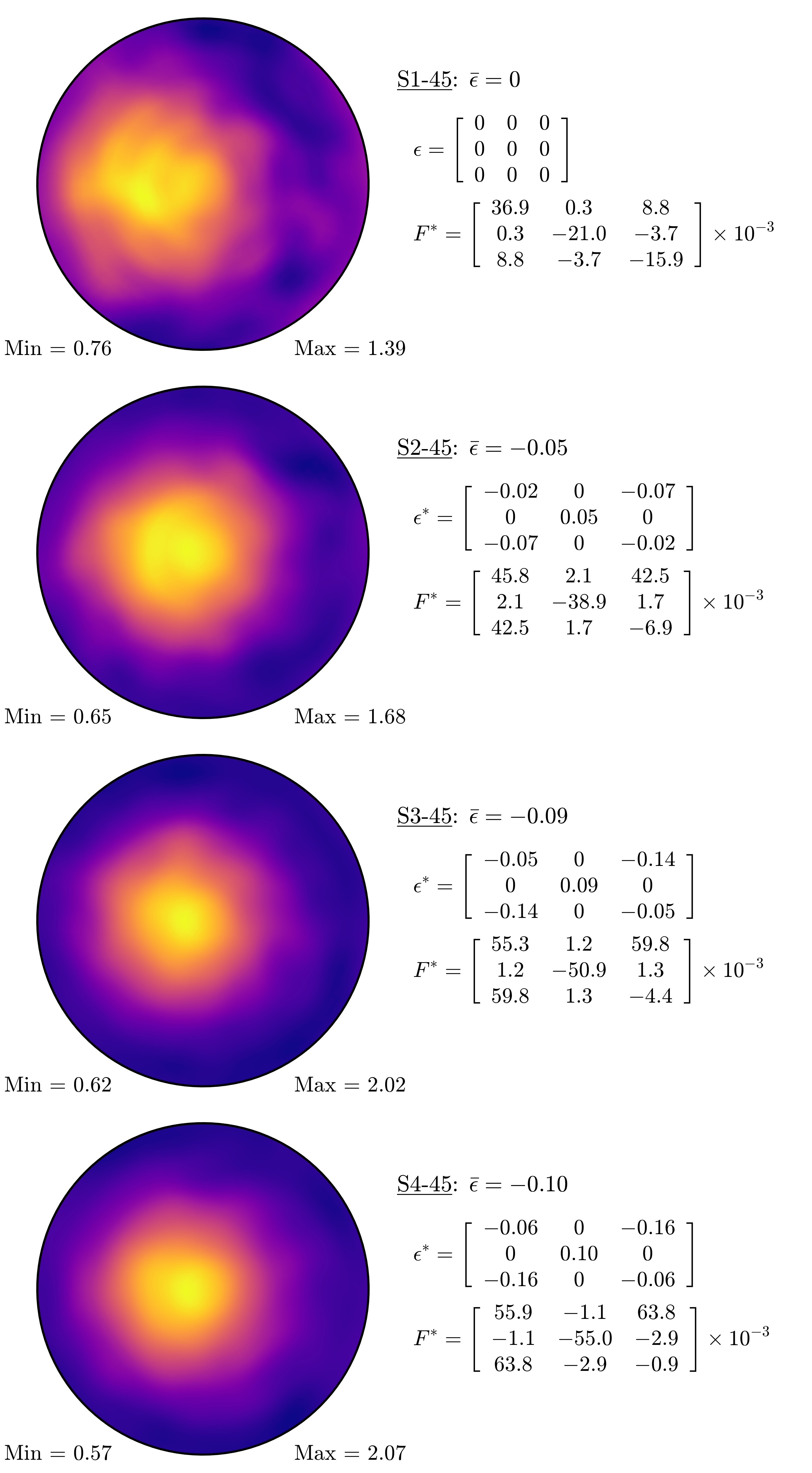}
       	\put(-200,360){(b)}\\
       	\includegraphics[width=0.35\linewidth]{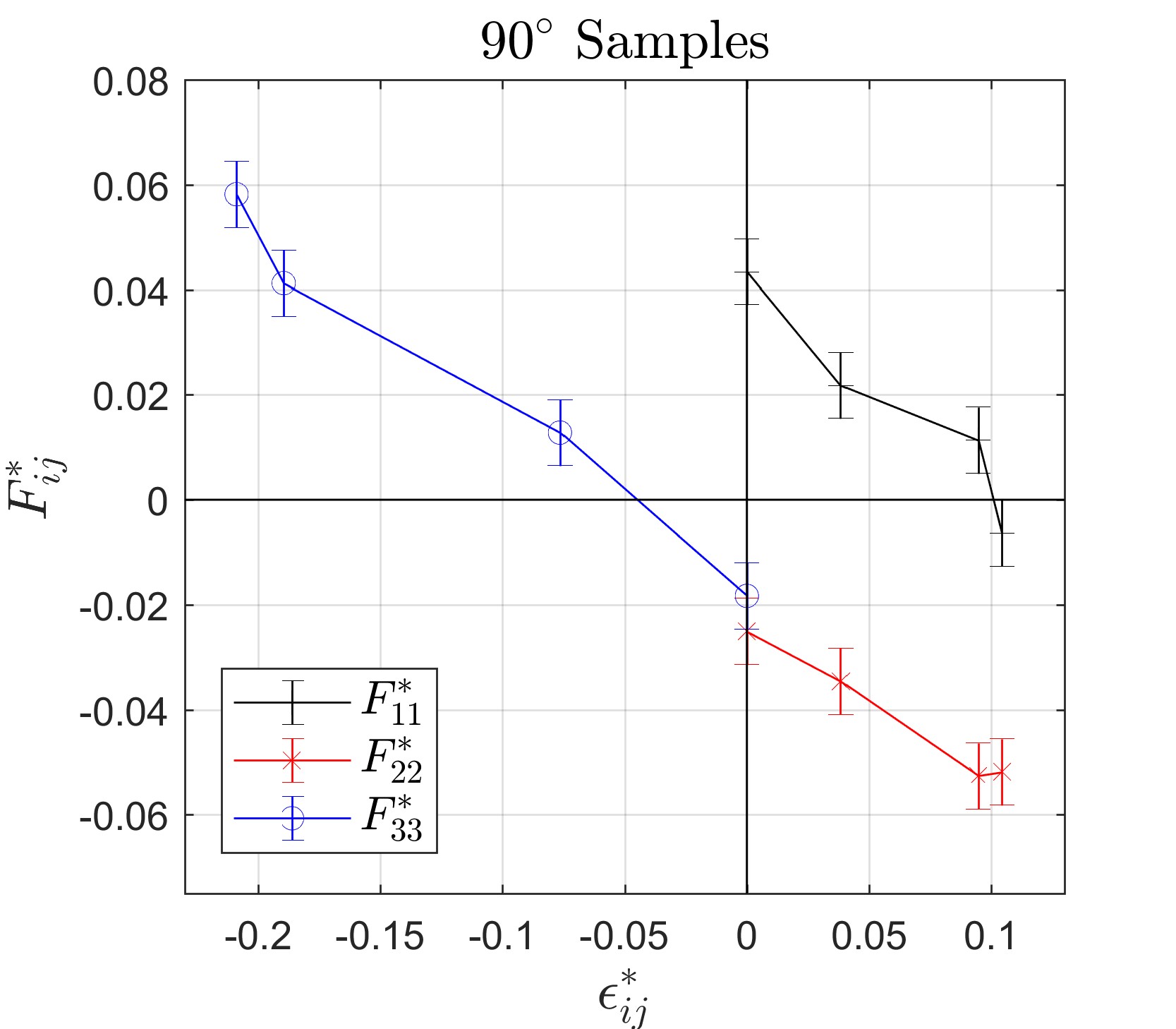}
       	\put(-190,145){(c)}
       	\includegraphics[width=0.35\linewidth]{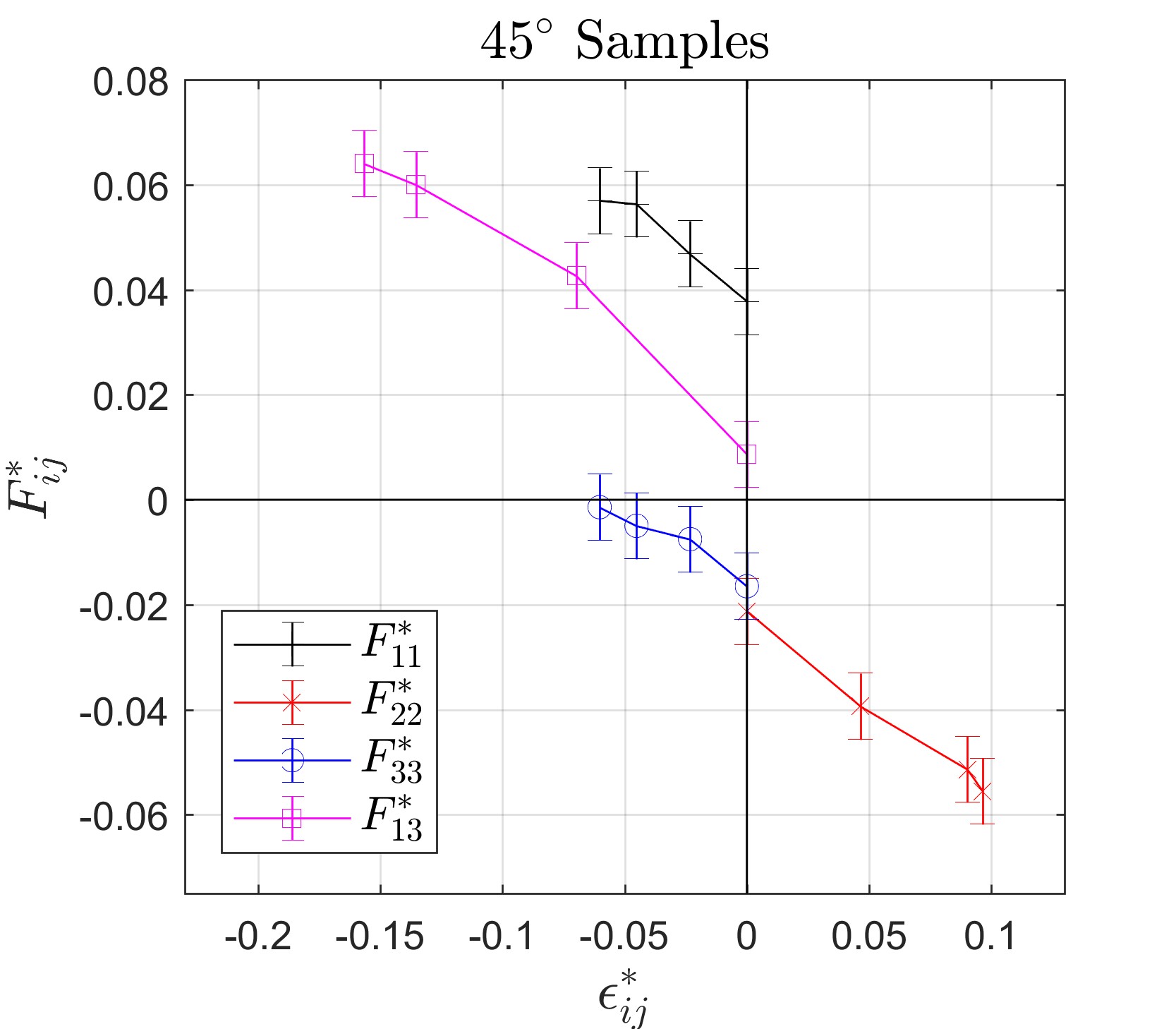}
       	\put(-190,145){(d)}\\
	\caption{(colour online) Evolution of structure within clay samples as a function of strain.  With reference to Figure \ref{fig:ExpSetup}b, samples on the left are subject to strain at 90$^\circ$ to the initial compaction, samples on the right are at 45$^\circ$.  (a) and (b): pole figures show the three-dimensional orientation density as viewed in the direction of the second stage of compaction.  (c) and (d): individual components of the deviatoric fabric tensor as a function of the corresponding components of deviatoric strain.}
    	\label{fig:Exp2}
    	\centering
\end{figure*}

A large kaolin sample was compacted at 10\% moisture content within a 90mm diameter cylindrical mould to a density of 1.3g/cc.  This compaction was performed in layers to reduce density variation in the axial direction.  The initial state provided by this process approximately corresponded to sample S4-0 in the previous experiment.  A number of 15mm diameter cylindrical sub-samples were then cut from this larger sample; 4 samples with their axis at 90$^\circ$ to the initial compaction direction, 4 at 45$^\circ$ (see Figure \ref{fig:ExpSetup}c).  Each sub-sample was then compacted in the axial direction to final target densities summarised in Table \ref{tab:samples2}.  The initial length, and strain imposed by this second stage of compaction is also given in Table \ref{tab:samples2}.  As before, each of these samples was freeze-dried to remove all moisture prior to measurement.

 \begin{table}
 \caption{\label{tab:samples2} Sample properties for the second experiment.  All samples had 10\% moisture content }
 \begin{ruledtabular}
 \begin{tabular}{ c | c c c c c }
 
 Sample & Initial Length & Axial & Dry Density & Direction of\\
 Number & [mm] &  Strain &  [g/cc] & Compaction\\
 \hline
 S1-90 & 10 &  0 &  1.30 & 90$^\circ$\\
 S2-90 & 12.2 & 0.11 & 1.50 & 90$^\circ$\\
 S3-90 & 15.0 & 0.28 & 1.76 & 90$^\circ$\\
 S4-90 & 16.6 & 0.31 & 1.84 & 90$^\circ$\\
 \hline
 S1-45 & 10 & 0 &  1.30 & 45$^\circ$\\
 S2-45 & 12.2 &  0.11 & 1.51 & 45$^\circ$\\
 S3-45 & 15.0 &  0.27 & 1.76  & 45$^\circ$\\
 S4-45 & 16.6 &  0.29 & 1.84 & 45$^\circ$\\
 
 \end{tabular}
 \end{ruledtabular}
 \end{table}
 
Using the same instrument setup as the first experiment, the orientation density of these 8 samples was measured.  Figure \ref{fig:Exp2} shows the resulting pole-figures for each sample as viewed along the direction of the second stage of compaction.  Colourmaps are individually normalised with the maximum and minimum values indicated.  Shown alongside is the corresponding deviatoric strain ($\epsilon^*_{ij}=\epsilon_{ij}-\bar\epsilon \delta_{ij}$, where $\bar\epsilon=\tfrac{1}{3}\epsilon_{kk}$) and deviatoric fabric tensors, both expressed in the coordinate system indicated in Figure \ref{fig:ExpSetup}(c).  In Figure \ref{fig:Exp2}c and \ref{fig:Exp2}d, individual components of deviatoric fabric are plotted against corresponding components of deviatoric strain; components that do not significantly change are not shown.

For samples S1-90 to S4-90, we see that the second stage of compaction produces a significant transformation in the orientation density.  The initial state shows alignment in the $x$-direction in-line with the initial compaction.  As compressive strain is applied along the $z$-axis, the peak migrates and then intensifies in this direction.  This effect can also be observed in the deviatoric fabric which is approximately in principal directions over the whole loading path.  The diagonal terms are initially dominated by the $F^*_{11}$ component, which is offset by the $F^*_{22}$ and  $F^*_{33}$ components in roughly equal amounts.  As compressive strain is applied in the $z$-direction, the $F^*_{33}$ component grows at the expense of the other two.

In terms of orientation density, S1-45 to S4-45 show similar behaviour.  Over the second stage of compaction, the initial alignment of platelets transforms to align and intensify in the direction of compressive strain.  However, in contrast to the 90$^\circ$ samples, the final state is subtly misaligned; the peak value of the orientation density is slightly to the left of centre.  This is also observed in the fabric tensor which features a prominent $F^*_{13}$ component in the final state.  Note that this departure from principal directions is also present in the applied strain.

When viewed in relation to individual components of the strain tensor, the evolution of the fabric tensor becomes clear.  From Figure \ref{fig:Exp2}c and \ref{fig:Exp2}d we see that, within experimental error, components of the deviatoric fabric tensor change from their initial state at a constant rate with respect to the corresponding component of deviatoric strain.  Of particular note; this rate of change appears consistent across all components.  Averaging across components provides the following simple relationship;
\begin{equation}
\frac{\partial F^*_{ij}}{\partial \epsilon^*_{ij}}=-0.35 \pm 0.01
\end{equation}

This constant relationship suggests that isotropic strain paths have no impact on fabric, and that principal directions of strain and fabric should coincide if the initial state is random.  However, it should be made clear that this relationship was only observed to hold over the range and direction of deformations examined in the second experiment.  Outside of this range, or in the case of other strain paths (e.g. pure shear or dilation), the situation may vary.  In fact, a similar relationship can be observed in the first experiment from samples S1-0 to S4-0, but with a different constant of proportionality ($-0.24$ rather than $-0.35$).  This was a curious result that prompted further examination, including measurements from additional samples uniaxially compacted from an initial state corresponding to sample S6-0.  All of this additional work confirmed the original observation.  Work continues on resolving this detail, however outside of this small issue, the results suggest that a relatively simple relationship between strain and fabric is available.  

In conclusion, neutron diffraction techniques provide a direct approach for assessing fabric within clay samples.  Through this approach we have been able to directly observe the orientation density of clay platelets and uncover a simple empirical relationship between deviatoric strain and the evolution of the fabric tensor.  Given the direct link between mechanical properties and fabric, it is conceivable that this simple relationship could be integrated into an anisotropic constitutive model for this type of material.  As opposed to current phenomenological approaches, this may allow the development of new anisotropic constitutive models based directly on the micromechanics of clay materials.

This work was made possible by the Australian Centre for Neutron Scattering (ACNS) through their user access program (proposals 4105 and 4308).  Additional support was also provided by the Australian Institute of Nuclear Science and Engineering (AINSE).

\bibliography{refs}

\begin{thebibliography}{19}%
\makeatletter
\providecommand \@ifxundefined [1]{%
 \@ifx{#1\undefined}
}%
\providecommand \@ifnum [1]{%
 \ifnum #1\expandafter \@firstoftwo
 \else \expandafter \@secondoftwo
 \fi
}%
\providecommand \@ifx [1]{%
 \ifx #1\expandafter \@firstoftwo
 \else \expandafter \@secondoftwo
 \fi
}%
\providecommand \natexlab [1]{#1}%
\providecommand \enquote  [1]{``#1''}%
\providecommand \bibnamefont  [1]{#1}%
\providecommand \bibfnamefont [1]{#1}%
\providecommand \citenamefont [1]{#1}%
\providecommand \href@noop [0]{\@secondoftwo}%
\providecommand \href [0]{\begingroup \@sanitize@url \@href}%
\providecommand \@href[1]{\@@startlink{#1}\@@href}%
\providecommand \@@href[1]{\endgroup#1\@@endlink}%
\providecommand \@sanitize@url [0]{\catcode `\\12\catcode `\$12\catcode
  `\&12\catcode `\#12\catcode `\^12\catcode `\_12\catcode `\%12\relax}%
\providecommand \@@startlink[1]{}%
\providecommand \@@endlink[0]{}%
\providecommand \url  [0]{\begingroup\@sanitize@url \@url }%
\providecommand \@url [1]{\endgroup\@href {#1}{\urlprefix }}%
\providecommand \urlprefix  [0]{URL }%
\providecommand \Eprint [0]{\href }%
\providecommand \doibase [0]{http://dx.doi.org/}%
\providecommand \selectlanguage [0]{\@gobble}%
\providecommand \bibinfo  [0]{\@secondoftwo}%
\providecommand \bibfield  [0]{\@secondoftwo}%
\providecommand \translation [1]{[#1]}%
\providecommand \BibitemOpen [0]{}%
\providecommand \bibitemStop [0]{}%
\providecommand \bibitemNoStop [0]{.\EOS\space}%
\providecommand \EOS [0]{\spacefactor3000\relax}%
\providecommand \BibitemShut  [1]{\csname bibitem#1\endcsname}%
\let\auto@bib@innerbib\@empty
\bibitem [{\citenamefont {Wheeler}\ \emph {et~al.}(2003)\citenamefont
  {Wheeler}, \citenamefont {N{\"a}{\"a}t{\"a}nen}, \citenamefont {Karstunen},\
  and\ \citenamefont {Lojander}}]{wheeler2003anisotropic}%
  \BibitemOpen
  \bibfield  {author} {\bibinfo {author} {\bibfnamefont {S.~J.}\ \bibnamefont
  {Wheeler}}, \bibinfo {author} {\bibfnamefont {A.}~\bibnamefont
  {N{\"a}{\"a}t{\"a}nen}}, \bibinfo {author} {\bibfnamefont {M.}~\bibnamefont
  {Karstunen}}, \ and\ \bibinfo {author} {\bibfnamefont {M.}~\bibnamefont
  {Lojander}},\ }\href@noop {} {\bibfield  {journal} {\bibinfo  {journal}
  {Canadian Geotechnical Journal}\ }\textbf {\bibinfo {volume} {40}},\ \bibinfo
  {pages} {403} (\bibinfo {year} {2003})}\BibitemShut {NoStop}%
\bibitem [{\citenamefont {Voyiadjis}\ and\ \citenamefont
  {Song}(2000)}]{voyiadjis2000finite}%
  \BibitemOpen
  \bibfield  {author} {\bibinfo {author} {\bibfnamefont {G.~Z.}\ \bibnamefont
  {Voyiadjis}}\ and\ \bibinfo {author} {\bibfnamefont {C.~R.}\ \bibnamefont
  {Song}},\ }\href@noop {} {\bibfield  {journal} {\bibinfo  {journal} {Journal
  of engineering mechanics}\ }\textbf {\bibinfo {volume} {126}},\ \bibinfo
  {pages} {1012} (\bibinfo {year} {2000})}\BibitemShut {NoStop}%
\bibitem [{\citenamefont {Barden}\ \emph {et~al.}(1973)\citenamefont {Barden},
  \citenamefont {McGown},\ and\ \citenamefont {Collins}}]{barden1973collapse}%
  \BibitemOpen
  \bibfield  {author} {\bibinfo {author} {\bibfnamefont {L.}~\bibnamefont
  {Barden}}, \bibinfo {author} {\bibfnamefont {A.}~\bibnamefont {McGown}}, \
  and\ \bibinfo {author} {\bibfnamefont {K.}~\bibnamefont {Collins}},\
  }\href@noop {} {\bibfield  {journal} {\bibinfo  {journal} {Engineering
  Geology}\ }\textbf {\bibinfo {volume} {7}},\ \bibinfo {pages} {49} (\bibinfo
  {year} {1973})}\BibitemShut {NoStop}%
\bibitem [{\citenamefont {Collins}\ and\ \citenamefont
  {McGown}(1974)}]{collins1974form}%
  \BibitemOpen
  \bibfield  {author} {\bibinfo {author} {\bibfnamefont {K.~T.}\ \bibnamefont
  {Collins}}\ and\ \bibinfo {author} {\bibfnamefont {A.}~\bibnamefont
  {McGown}},\ }\href@noop {} {\bibfield  {journal} {\bibinfo  {journal}
  {Geotechnique}\ }\textbf {\bibinfo {volume} {24}},\ \bibinfo {pages} {223}
  (\bibinfo {year} {1974})}\BibitemShut {NoStop}%
\bibitem [{\citenamefont {Osipov}\ and\ \citenamefont
  {Sokolov}(1978)}]{osipov1978structure}%
  \BibitemOpen
  \bibfield  {author} {\bibinfo {author} {\bibfnamefont {V.}~\bibnamefont
  {Osipov}}\ and\ \bibinfo {author} {\bibfnamefont {V.}~\bibnamefont
  {Sokolov}},\ }\href@noop {} {\bibfield  {journal} {\bibinfo  {journal}
  {Bulletin of Engineering Geology and the Environment}\ }\textbf {\bibinfo
  {volume} {18}},\ \bibinfo {pages} {83} (\bibinfo {year} {1978})}\BibitemShut
  {NoStop}%
\bibitem [{\citenamefont {Sloane}\ and\ \citenamefont
  {Kell}(1966)}]{sloane1966fabric}%
  \BibitemOpen
  \bibfield  {author} {\bibinfo {author} {\bibfnamefont {R.~L.}\ \bibnamefont
  {Sloane}}\ and\ \bibinfo {author} {\bibfnamefont {T.}~\bibnamefont {Kell}},\
  }in\ \href@noop {} {\emph {\bibinfo {booktitle} {Clays and Clay Minerals:
  Proceedings of the Fourteenth National Conference, Berkeley, California}}}\
  (\bibinfo {organization} {Elsevier},\ \bibinfo {year} {1966})\ p.\ \bibinfo
  {pages} {289}\BibitemShut {NoStop}%
\bibitem [{\citenamefont {Diamond}(1970)}]{diamond1970pore}%
  \BibitemOpen
  \bibfield  {author} {\bibinfo {author} {\bibfnamefont {S.}~\bibnamefont
  {Diamond}},\ }\href@noop {} {\bibfield  {journal} {\bibinfo  {journal} {Clays
  and clay minerals}\ }\textbf {\bibinfo {volume} {18}},\ \bibinfo {pages} {7}
  (\bibinfo {year} {1970})}\BibitemShut {NoStop}%
\bibitem [{\citenamefont {Romero}\ and\ \citenamefont
  {Simms}(2008)}]{romero2008microstructure}%
  \BibitemOpen
  \bibfield  {author} {\bibinfo {author} {\bibfnamefont {E.}~\bibnamefont
  {Romero}}\ and\ \bibinfo {author} {\bibfnamefont {P.~H.}\ \bibnamefont
  {Simms}},\ }in\ \href@noop {} {\emph {\bibinfo {booktitle} {Laboratory and
  Field Testing of Unsaturated Soils}}}\ (\bibinfo  {publisher} {Springer},\
  \bibinfo {year} {2008})\ pp.\ \bibinfo {pages} {93--115}\BibitemShut
  {NoStop}%
\bibitem [{\citenamefont {Montes-H}\ \emph {et~al.}(2005)\citenamefont
  {Montes-H}, \citenamefont {Geraud}, \citenamefont {Duplay},\ and\
  \citenamefont {Reuschle}}]{montes2005esem}%
  \BibitemOpen
  \bibfield  {author} {\bibinfo {author} {\bibfnamefont {G.}~\bibnamefont
  {Montes-H}}, \bibinfo {author} {\bibfnamefont {Y.}~\bibnamefont {Geraud}},
  \bibinfo {author} {\bibfnamefont {J.}~\bibnamefont {Duplay}}, \ and\ \bibinfo
  {author} {\bibfnamefont {T.}~\bibnamefont {Reuschle}},\ }\href@noop {}
  {\bibfield  {journal} {\bibinfo  {journal} {Colloids and Surfaces A:
  Physicochemical and Engineering Aspects}\ }\textbf {\bibinfo {volume}
  {262}},\ \bibinfo {pages} {14} (\bibinfo {year} {2005})}\BibitemShut
  {NoStop}%
\bibitem [{\citenamefont {Lin}\ and\ \citenamefont
  {Cerato}(2014)}]{lin2014applications}%
  \BibitemOpen
  \bibfield  {author} {\bibinfo {author} {\bibfnamefont {B.}~\bibnamefont
  {Lin}}\ and\ \bibinfo {author} {\bibfnamefont {A.~B.}\ \bibnamefont
  {Cerato}},\ }\href@noop {} {\bibfield  {journal} {\bibinfo  {journal}
  {Engineering Geology}\ }\textbf {\bibinfo {volume} {177}},\ \bibinfo {pages}
  {66} (\bibinfo {year} {2014})}\BibitemShut {NoStop}%
\bibitem [{\citenamefont {Suuronen}\ \emph {et~al.}(2014)\citenamefont
  {Suuronen}, \citenamefont {Matusewicz}, \citenamefont {Olin},\ and\
  \citenamefont {Serimaa}}]{suuronen2014x}%
  \BibitemOpen
  \bibfield  {author} {\bibinfo {author} {\bibfnamefont {J.-P.}\ \bibnamefont
  {Suuronen}}, \bibinfo {author} {\bibfnamefont {M.}~\bibnamefont
  {Matusewicz}}, \bibinfo {author} {\bibfnamefont {M.}~\bibnamefont {Olin}}, \
  and\ \bibinfo {author} {\bibfnamefont {R.}~\bibnamefont {Serimaa}},\
  }\href@noop {} {\bibfield  {journal} {\bibinfo  {journal} {Applied Clay
  Science}\ }\textbf {\bibinfo {volume} {101}},\ \bibinfo {pages} {401}
  (\bibinfo {year} {2014})}\BibitemShut {NoStop}%
\bibitem [{\citenamefont {Haines}\ \emph {et~al.}(2009)\citenamefont {Haines},
  \citenamefont {van~der Pluijm}, \citenamefont {Ikari}, \citenamefont
  {Saffer},\ and\ \citenamefont {Marone}}]{haines2009clay}%
  \BibitemOpen
  \bibfield  {author} {\bibinfo {author} {\bibfnamefont {S.~H.}\ \bibnamefont
  {Haines}}, \bibinfo {author} {\bibfnamefont {B.~A.}\ \bibnamefont {van~der
  Pluijm}}, \bibinfo {author} {\bibfnamefont {M.~J.}\ \bibnamefont {Ikari}},
  \bibinfo {author} {\bibfnamefont {D.~M.}\ \bibnamefont {Saffer}}, \ and\
  \bibinfo {author} {\bibfnamefont {C.}~\bibnamefont {Marone}},\ }\href@noop {}
  {\bibfield  {journal} {\bibinfo  {journal} {Journal of Geophysical Research:
  Solid Earth}\ }\textbf {\bibinfo {volume} {114}} (\bibinfo {year}
  {2009})}\BibitemShut {NoStop}%
\bibitem [{\citenamefont {Schumann}\ \emph {et~al.}(2014)\citenamefont
  {Schumann}, \citenamefont {Stipp}, \citenamefont {Leiss},\ and\ \citenamefont
  {Behrmann}}]{schumann2014texture}%
  \BibitemOpen
  \bibfield  {author} {\bibinfo {author} {\bibfnamefont {K.}~\bibnamefont
  {Schumann}}, \bibinfo {author} {\bibfnamefont {M.}~\bibnamefont {Stipp}},
  \bibinfo {author} {\bibfnamefont {B.}~\bibnamefont {Leiss}}, \ and\ \bibinfo
  {author} {\bibfnamefont {J.~H.}\ \bibnamefont {Behrmann}},\ }\href@noop {}
  {\bibfield  {journal} {\bibinfo  {journal} {Tectonophysics}\ }\textbf
  {\bibinfo {volume} {636}},\ \bibinfo {pages} {125} (\bibinfo {year}
  {2014})}\BibitemShut {NoStop}%
\bibitem [{\citenamefont {Voltolini}\ \emph {et~al.}(2008)\citenamefont
  {Voltolini}, \citenamefont {Wenk}, \citenamefont {Mondol}, \citenamefont
  {Bj{\o}rlykke},\ and\ \citenamefont {Jahren}}]{voltolini2008anisotropy}%
  \BibitemOpen
  \bibfield  {author} {\bibinfo {author} {\bibfnamefont {M.}~\bibnamefont
  {Voltolini}}, \bibinfo {author} {\bibfnamefont {H.-R.}\ \bibnamefont {Wenk}},
  \bibinfo {author} {\bibfnamefont {N.~H.}\ \bibnamefont {Mondol}}, \bibinfo
  {author} {\bibfnamefont {K.}~\bibnamefont {Bj{\o}rlykke}}, \ and\ \bibinfo
  {author} {\bibfnamefont {J.}~\bibnamefont {Jahren}},\ }\href@noop {}
  {\bibfield  {journal} {\bibinfo  {journal} {Geophysics}\ }\textbf {\bibinfo
  {volume} {74}},\ \bibinfo {pages} {D13} (\bibinfo {year} {2008})}\BibitemShut
  {NoStop}%
\bibitem [{\citenamefont {Cifelli}\ \emph {et~al.}(2005)\citenamefont
  {Cifelli}, \citenamefont {Mattei}, \citenamefont {Chadima}, \citenamefont
  {Hirt},\ and\ \citenamefont {Hansen}}]{cifelli2005origin}%
  \BibitemOpen
  \bibfield  {author} {\bibinfo {author} {\bibfnamefont {F.}~\bibnamefont
  {Cifelli}}, \bibinfo {author} {\bibfnamefont {M.}~\bibnamefont {Mattei}},
  \bibinfo {author} {\bibfnamefont {M.}~\bibnamefont {Chadima}}, \bibinfo
  {author} {\bibfnamefont {A.}~\bibnamefont {Hirt}}, \ and\ \bibinfo {author}
  {\bibfnamefont {A.}~\bibnamefont {Hansen}},\ }\href@noop {} {\bibfield
  {journal} {\bibinfo  {journal} {Earth and Planetary Science Letters}\
  }\textbf {\bibinfo {volume} {235}},\ \bibinfo {pages} {62} (\bibinfo {year}
  {2005})}\BibitemShut {NoStop}%
\bibitem [{\citenamefont {Wenk}(2006)}]{wenk2006neutron}%
  \BibitemOpen
  \bibfield  {author} {\bibinfo {author} {\bibfnamefont {H.-R.}\ \bibnamefont
  {Wenk}},\ }\href@noop {} {\bibfield  {journal} {\bibinfo  {journal} {Reviews
  in Mineralogy and Geochemistry}\ }\textbf {\bibinfo {volume} {63}},\ \bibinfo
  {pages} {399} (\bibinfo {year} {2006})}\BibitemShut {NoStop}%
\bibitem [{\citenamefont {Ken-Ichi}(1984)}]{ken1984distribution}%
  \BibitemOpen
  \bibfield  {author} {\bibinfo {author} {\bibfnamefont {K.}~\bibnamefont
  {Ken-Ichi}},\ }\href@noop {} {\bibfield  {journal} {\bibinfo  {journal}
  {International Journal of Engineering Science}\ }\textbf {\bibinfo {volume}
  {22}},\ \bibinfo {pages} {149} (\bibinfo {year} {1984})}\BibitemShut
  {NoStop}%
\bibitem [{\citenamefont {Delage}\ \emph {et~al.}(1982)\citenamefont {Delage},
  \citenamefont {Tessier},\ and\ \citenamefont
  {Marcel-Audiguier}}]{delage1982use}%
  \BibitemOpen
  \bibfield  {author} {\bibinfo {author} {\bibfnamefont {P.}~\bibnamefont
  {Delage}}, \bibinfo {author} {\bibfnamefont {D.}~\bibnamefont {Tessier}}, \
  and\ \bibinfo {author} {\bibfnamefont {M.}~\bibnamefont {Marcel-Audiguier}},\
  }\href@noop {} {\bibfield  {journal} {\bibinfo  {journal} {Canadian
  Geotechnical Journal}\ }\textbf {\bibinfo {volume} {19}},\ \bibinfo {pages}
  {111} (\bibinfo {year} {1982})}\BibitemShut {NoStop}%
\bibitem [{\citenamefont {Burton}\ \emph {et~al.}(2015)\citenamefont {Burton},
  \citenamefont {Pineda}, \citenamefont {Sheng},\ and\ \citenamefont
  {Airey}}]{burton2015microstructural}%
  \BibitemOpen
  \bibfield  {author} {\bibinfo {author} {\bibfnamefont {G.~J.}\ \bibnamefont
  {Burton}}, \bibinfo {author} {\bibfnamefont {J.~A.}\ \bibnamefont {Pineda}},
  \bibinfo {author} {\bibfnamefont {D.}~\bibnamefont {Sheng}}, \ and\ \bibinfo
  {author} {\bibfnamefont {D.}~\bibnamefont {Airey}},\ }\href@noop {}
  {\bibfield  {journal} {\bibinfo  {journal} {Engineering Geology}\ }\textbf
  {\bibinfo {volume} {193}},\ \bibinfo {pages} {363} (\bibinfo {year}
  {2015})}\BibitemShut {NoStop}%
\end{thebibliography}%

\end{document}